\documentclass[sigconf,dvipsnames]{acmart}
\acmConference[ASE 2022]{37th IEEE/ACM International Conference on Automated Software Engineering}{Mon 10 - Fri 14 October 2022}{Ann Arbor, Michigan, United States}
\usepackage{listings}
\usepackage{color}
\usepackage{xspace}
\usepackage{xcolor}
\usepackage{graphicx}
\usepackage{stfloats}
\usepackage{enumitem}
\usepackage{amsmath}
\usepackage{fancybox}
\usepackage[ruled,lined,linesnumbered,vlined,algo2e]{algorithm2e}
\usepackage{soul}
\usepackage{url}
\usepackage{calc}
\usepackage{caption}
\makeatletter
\newcommand\xleftrightarrow[2][]{%
  \ext@arrow 9999{\longleftrightarrowfill@}{#1}{#2}}
\newcommand\longleftrightarrowfill@{%
  \arrowfill@\leftarrow\relbar\rightarrow}
\makeatother

\newcommand{\lstbg}[3][0pt]{{\fboxsep#1\colorbox{#2}{\strut #3}}}
\definecolor{codegreen}{rgb}{0,0.6,0}
\lstdefinelanguage{diff}{
	frame=single,
	basicstyle=\ttfamily\scriptsize\bfseries,
	morecomment=[f][\color{red}]{---}, 
	morecomment=[f][\color{codegreen}]{+++},
	morecomment=[f][\lstbg{red!20}]{-},
	morecomment=[f][\lstbg{green!20}]{+},
	morecomment=[f][\color{blue}]{@@},
}
\usepackage{microtype}
\setlist[itemize]{noitemsep, topsep=0pt}
\AtBeginDocument{%
  \providecommand\BibTeX{{%
    \normalfont B\kern-0.5em{\scshape i\kern-0.25em b}\kern-0.8em\TeX}}}

\newcommand{\tool}{\textsc{\textsc{Sembid}}\xspace}

\newcommand{\ly}[1][\textcolor{black}]{#1}%
\newcommand{\lyu}[1][\textcolor{black}]{#1}

\newcommand{\synb}{SynB\xspace}
\newcommand{\semb}{SemB\xspace}
\newcommand{\pnm}{\textit{Patch} and \textit{Minor}\xspace}
\newcommand{\pa}{\textit{Patch}\xspace}
\newcommand{\mi}{\textit{Minor}\xspace}
\newcommand{\ma}{\textit{Major}\xspace}
\newcommand{\subgraph}{cluster\xspace}
\newcommand{\subgraphs}{clusters\xspace}

\copyrightyear{2022}
\acmYear{2022}
\setcopyright{acmcopyright}\acmConference[ASE '22]{37th IEEE/ACM International
Conference on Automated Software Engineering}{October 10--14, 2022}{Rochester,
MI, USA}
\acmBooktitle{37th IEEE/ACM International Conference on Automated Software
Engineering (ASE '22), October 10--14, 2022, Rochester, MI, USA}
\acmPrice{15.00}
\acmDOI{10.1145/3551349.3556956}
\acmISBN{978-1-4503-9475-8/22/10}

\begin{document}
\abovedisplayskip=6pt
\abovedisplayshortskip=6pt
\belowdisplayskip=6pt
\belowdisplayshortskip=6pt
\title[Has My Release Disobeyed Semantic Versioning? Static Detection Based on Semantic Differencing]{Has My Release Disobeyed Semantic Versioning? Static Detection Based on Semantic Differencing}
\author{Lyuye Zhang}
\orcid{0000-0003-3087-9645}
\affiliation{%
  \institution{School of Computer Science and Engineering, Nanyang Technological University}
    \city{Singapore}
  \country{Singapore}
}
\email{zh0004ye@e.ntu.edu.sg}

\author{Chengwei Liu}
\orcid{0000-0003-1175-2753}
\affiliation{%
  \institution{School of Computer Science and Engineering, Nanyang Technological University}
    \city{Singapore}
  \country{Singapore}
}
\email{chengwei001@e.ntu.edu.sg}

\author{Zhengzi Xu}
\orcid{0000-0002-8390-7518}

\authornote{Zhengzi Xu is the corresponding author.}
\affiliation{%
  \institution{School of Computer Science and Engineering, Nanyang Technological University}
    \city{Singapore}
  \country{Singapore}
}
\email{zhengzi.xu@ntu.edu.sg}

\author{Sen Chen}
\orcid{0000-0001-9477-4100}

\affiliation{%
  \institution{College of Intelligence and Computing, Tianjin University}
  \city{Tianjin}
  \country{China}}
\email{senchen@tju.edu.cn}

\author{Lingling Fan}
\orcid{0000-0002-2428-9297}
\affiliation{
  \institution{College of Cyber Science, Nankai University}
  \city{Tianjin}
  \country{China}
}
\email{linglingfan@nankai.edu.cn}

\author{Bihuan Chen}
\affiliation{%
 \institution{School of Computer Science and
Shanghai Key Laboratory of Data
Science, Fudan University}
 \city{Shanghai}
 \country{China}}
\email{bhchen@fudan.edu.cn}

\author{Yang Liu}
\orcid{0000-0001-7300-9215}
\affiliation{%
  \institution{School of Computer Science and Engineering, Nanyang Technological University}
  \city{Singapore}
  \country{Singapore}}
\email{yangliu@ntu.edu.sg}
\begin{abstract}
To enhance the compatibility in the version control of Java Third-party Libraries (TPLs), Maven adopts Semantic Versioning (SemVer) to standardize the underlying meaning of versions, but users could still confront abnormal execution and crash after upgrades even if compilation and linkage succeed. It is caused by semantic breaking (\semb) issues, such that APIs directly used by users have identical signatures but inconsistent semantics across upgrades. To strengthen compliance with SemVer rules, developers and users should be alerted of such issues. Unfortunately, it is challenging to detect them statically, because semantic changes in the internal methods of APIs are difficult to capture. Dynamic testing can confirmingly uncover some, but it is limited by inadequate coverage.

To detect \semb issues over compatible upgrades (\pnm) by SemVer rules, we conduct an empirical study on 180 \semb issues to understand the root causes, inspired by which, we propose \tool (Semantic Breaking Issue Detector) to statically detect such issues of TPLs for developers and users. Since APIs are directly used by users, \tool detects and reports \semb issues based on APIs.
For a pair of APIs, \tool walks through the call chains originating from the API to locate breaking changes by measuring semantic diff.
Then, \tool checks if the breaking changes can affect API's output along call chains.
The evaluation showed \tool achieved $90.26\%$ recall and $81.29\%$ precision and outperformed other API checkers on \semb API detection. We also revealed \tool detected over 3 times more \semb APIs with better coverage than unit tests, the commonly used solution.
Furthermore, we carried out an empirical study
on $1,629,589$ APIs
from $546$ version pairs of top Java libraries and found there were 2$\sim$4 times more \semb APIs than those with signature-based issues. Due to various version release strategies, $33.83\%$ of \pa version pairs and $64.42\%$ of \mi version pairs had at least one API affected by any breaking.
\end{abstract}
\maketitle

\section{Introduction}
\label{introduction}

The frequent updates of Third-party Libraries (TPLs) prompt downstream users to upgrade their dependencies accordingly to embrace necessary new features and bug fixes \cite{zhan2021atvhunter,zhan2021research}, while the potentially incompatible changes could lead to abnormal execution or even crashes of downstream projects according to \cite{mostafa2017experience}.
To address this problem, modern package managers adopt Semantic Versioning (SemVer)~\cite{semver} to control the compatible upgrading of TPLs. SemVer divides upgrades into three types, namely, \textit{Major}, \textit{Minor}, and \textit{Patch} and requires changes in \textit{Minor} and \textit{Patch} upgrades to be backward compatible. Unfortunately, it is unlikely to guarantee all versions strictly satisfy SemVer rules, especially in legacy platforms like Maven~\cite{maven} for Java.
For example, \textit{Hadoop-hdfs} \cite{hadoop}, a widely-used framework of Apache was susceptible to a breaking issue \cite{hdfs} during \textit{Patch} upgrade ($3.0.0$-$3.0.1$). 
Not limited to popular Java projects like \textit{Hadoop}, the breaking issues commonly exist in
the Maven ecosystem.
As revealed by
Ochoa et al.~\cite{ochoa2021breaking},
$20.1\%$ of non-\textit{Major} upgrades in the Maven repository~\cite{mvnrepo} have introduced breaking changes that could induce massive unexpected downstream issues.

In recent years, works have been proposed to detect the breaking compatibility issues at the API level of TPLs.
Since APIs serve as the entry points for downstream projects to access TPLs, an upgrade is usually considered compatible if all APIs are compatible.
Many API compatibility checkers~\cite{jour, japitools, jchecker, sigtest, revapi, japi-compliance-checker} are proposed to uncover signature-based errors, such as \textit{ClassNotFoundException}
and \textit{NoSuchMethodError}. 
We call these errors \textit{Syntactic Breaking} (\synb) in this paper.
However, these checkers fail to detect the breaking issues caused by the internal behavioral changes, which are often exposed at run-time, such as abnormal output, unexpected exceptions, and even crashes \cite{fan2018large,su2020my}. That is because internal changes often dwell in the bodies of indirectly called methods that are not considered by these checkers. 
We call the internal behavioral changes without syntactic errors \textit{Semantic Breaking} (\semb).

Unlike \synb, \semb is much more difficult to be detected. Existing works mostly rely on manually curated dynamic regression tests to detect them. For example, 
Mostafa et al.~\cite{mostafa2017experience} found only 13 of 126 real-world \semb APIs could be detected by existing unit tests. 
Chen et al.~\cite{chen2020taming} have proposed DeBBI which accelerates unit tests from downstream projects to facilitate the detection, but it still relies on testing which is handicapped 
by the coverage of limited test cases. Therefore, detecting \semb issues properly is challenging, without which, SemVer rules cannot be completely enforced. Then, unexpected breaking upgrades would sabotage the community.

To bridge this gap, we seek to detect \semb statically for better coverage of potential issues. We are facing the following \textbf{challenges}. \textbf{C1}: It is unclear how to statically infer dynamic behavioral changes by source code diff, which has hardly been studied. 
\textbf{C2}: Syntactic changes (including inter-method changes), such as the refactoring, which hardly changes the semantics, are difficult to be excluded. \textbf{C3}: Some behavioral changes, such as bug fixing, are considered compatible, and thus allowed by SemVer. They are supposed to be ruled out, but there is no specific criterion to identify them. \textbf{C4}: A considerable amount of internal changes implicitly optimize the logic or performance, which hardly changes the output of APIs, thus not observable from the user side. Without the impact analysis of these changes, naive detection may lead to false positives.

To this end, we propose, to the best of our knowledge, \tool (Semantic Breaking Issue Detector), the first static tool to detect potential \semb over \pnm upgrades against SemVer rules based on APIs. 
First, to address the challenge \textbf{C1}, \lyu{we conducted a study on causes of \semb which indicated inconsistent behaviors were reflected by the diff of dependency relationships derived from static slicing \cite{weiser1984program}}.
Then, to address syntactic changes in \textbf{C2}, \tool extracts abstract semantic information free from syntactic changes. 
It starts with Intermediate Representation (IR) and traverses the methods within the call chains of a given API pair to form \subgraphs of changed methods to neutralize inter-method changes. Given a \subgraph, \tool recursively backward slices the global output inter-procedurally to derive semantic relationships between global input and output, which eliminate local syntactic changes. \tool heuristically constructs semantic graphs based on sliced statements and measures their semantic diff based on subgraph isomorphism to infer the potential \semb.
For \textbf{C3}, during the semantic diff measurement, benign changes are identified and excluded by pre-summarised patterns to avoid false alarms. 
Last, for \textbf{C4}, along call chains from the \subgraph, \tool verifies if the captured \semb can influence the API's output by checking its triggerability as well as whether it can propagate back to the API.

To evaluate the accuracy of \semb detection of \tool, we first experimented for the benchmark with other API checkers. Due to the lack of a benchmark dataset, we manually formed one consisting of 77 version pairs with $671$ APIs. \tool outperforms other state-of-the-art Java API compatibility checkers \cite{revapi,japi-compliance-checker,clirr,sigtest,jchecker} with $90.26\%$ recall and $81.29\%$ precision. Then, over all APIs of breaking \pa upgrades, we compared \tool with unit tests in terms of the effectiveness of breaking API detection. 
It demonstrates that \tool with better API coverage could detect over 4 times more \semb APIs than unit tests.
Furthermore, we carried out a study on $21$ top Java libraries with $546$ version pairs from Maven Repository \cite{mvnrepo} to evaluate the compliance with SemVer at the levels of API, version pair, and library respectively. 
We found that, compared with $0.38\%$ and $1.04\%$ of APIs affected by \synb for \pnm respectively, $1.10\%$ and $4.06\%$ of APIs additionally brought in \semb issues. For version pairs, due to various
upgrading strategies adopted by libraries, $33.83\%$ of \pa and $64.42\%$ of \mi upgrades have breaking pairs.
We conclude our main contributions as follows:

\begin{itemize}[leftmargin=9pt]
    \item We conducted a study to understand the root causes of \semb and summarized the patterns of benign semantic changes.
    \item We proposed the first static \semb issue detector (\tool) to detect potential \semb issues against SemVer rules.
    \item We built a benchmark dataset of APIs for \semb detection to facilitate further research, which is accessible on our website~\cite{dataset}.\footnote{Data set is accessible at https://sites.google.com/view/ase22semverdetection/homepage}
    \item We carried out a study on compliance with SemVer rules in the top 21 real-world Java libraries. We found version release strategies adopted by libraries varied greatly so that SemVer rules were not reliable for users. The prevalence of \semb proved the necessity of detecting \semb when upgrading TPLs.
\end{itemize}

\section{Background and Motivation}\label{sec:background}

\subsection{Semantic Versioning Rules}
\label{definition}

The Semantic Versioning \cite{semver} stipulates rules for three upgrades:

\begin{itemize}[leftmargin=9pt]
    \item \textbf{Patch version Z (x.y.Z)}: \textit{"MUST be incremented if only backward-compatible bug fixes are introduced. A bug fix is defined as an internal change that fixes incorrect behavior."}
    \item \textbf{Minor version Y (x.Y.z)}: \textit{"MUST be incremented if new, backward-compatible functionality is introduced to the public API. It MUST be incremented if any API is deprecated."}
    \item \textbf{Major version X (X.y.z)}: \textit{"MUST be incremented if any backward-incompatible changes are introduced to the public API. It MAY also include minor and patch level changes."}
\end{itemize}

\label{prestudy}
\begin{lstlisting}[language=diff,caption=A Motivating Example from httpcore:4.2-4.3,label={lst:exp1},numbers=left]

-if (this.contentDecoder != null && 
-   (this.session.getEventMask()&SelectionKey.OP_READ)>0) {
+if (this.contentDecoder != null) {
+   while ((this.session.getEventMask()&SelectionKey.OP_READ)>0) {
       handler.inputReady(this, this.contentDecoder);
       if (this.contentDecoder.isCompleted()){
            resetInput();
+           break;
+       if (!this.inbuf.hasData()) 
+           break;
}}}

\end{lstlisting}

The breaking changes are not allowed in \pnm upgrades.
Hence, to enhance compliance with SemVer rules, \tool aims at detecting breaking changes based on API over \pnm upgrades to alert developers and users.
Unlike \synb, \semb can hardly be detected by existing tools so \tool focuses on detecting \semb to bridge the gap. 

\subsection{Motivating Example}
The example in Listing~\ref{lst:exp1} is used to demonstrate our motivation.

The code was collected from a Jira issue \cite{motivation} of \textit{http-core}~\cite{http}. 
The upgrade caused the decoder to be stuck in an infinite loop when it reaches the end of the buffer if the input buffer contains a character `\textbackslash r' that is never consumed.
First, the change was not documented as a breaking, which suggests the breaking was unexpected. Second, the unit tests failed to uncover the case. Last, the \mi upgrade from version 4.2 to 4.3 intuitively indicated backward compatibility. 
Therefore, the judgment of the author is not always reliable so 
a static tool with better coverage is needed to detect potential breaking issues for \pnm upgrades.

\section{Empirical Study}
In this section, we study the causes of \semb and the code patterns of benign changes.
\subsection{Study of Root Causes of Semantic Breaking}
To detect \semb, we conducted a study regarding its root causes.
\subsubsection{\textbf{Causes of \semb}}
\label{cause}

Mostafa et al.  \cite{mostafa2017experience} studied the Behavior Backward Incompatibilities in Java software, which also refers to the breaking APIs beyond signatures like \semb. 
They categorized the immediate causes into three: usage
change ($32.77\%$), e.g. enable/disable poor input; better output ($55.74\%$), e.g. output format change; and other reasons($11.49\%$), e.g. internal structure changes.

Despite the immediate causes, Mostafa et al. \cite{mostafa2017experience} only focused on the user side of APIs but failed to dive into the internal code to locate the root causes. Since no other works have studied the root causes of \semb, we conducted one by analyzing the 
internal code changes of APIs that caused \semb with the aid of
existing static java analysis tools, BCEL \cite{bcel} and Soot \cite{vallee2010soot}. The study 
included
180 real-world \semb issues (126 from \cite{mostafa2017experience}, 54 collected by ourselves in recent 3 years) with commits. We found \semb had various forms which could hardly be summarised as patterns at the source code level, but they usually occurred along with the following changes:

\begin{itemize}[leftmargin=9pt]
\item ($73.33\%$) The changed execution logic. In Listing \ref{lst:exp1}, the conditions remained the same (L2 \& L4), but the \textit{if} statement was altered to a \textit{while} loop, 
which led to an infinite loop.
\item ($91.67\%$) The changed calculation of the output. For example, in a self-increment function, the change of calculation from $a += 1$ to $a += 2$ obviously modifies the output, while the execution logic remains the same.
\end{itemize}
The two types of causes overlapped in over $60\%$ of issues. The first change in the execution logic embodied the inconsistency in Control Flow Graph (CFG). The second change in the calculation process, embodied in the changed data flow, can reflect the inconsistency of output values. For $86\%$ of cases, the \semb changes dwelt in internal methods called by APIs instead of entry methods. Therefore, internal code analysis is required to detect \semb.

\subsubsection{\textbf{What Changes Will NOT Cause \semb?}}
\label{noise}

Since successful upgrades have no specific indicator (no-issue is not enough), we could only study successful regression tests to understand why some changes do not cause \semb. We firstly collected 20 most used Maven libraries according to Maven Repository \cite{mvnrepo} with 77 version pairs and ran regression tests by testing the new implementation with old unit tests. $20,373$ tested APIs were derived. Then, we
filtered out APIs with no changes in called methods to obtain
$2,191$ APIs. $500$ successful cases with binary change were randomly selected. \ly{For each case, we manually analyzed the diff dwelling in the methods called by the APIs to check if there existed an input to trigger the failure of unit tests. If the input did not exist, the reasons were collected.} The cases of each reason may overlap with one another. The APIs passed regression tests due to:

\begin{itemize}[leftmargin=9pt]
\item ($27.4\%$) \textbf{Inadequate input to trigger \semb.} We manually examined these cases and found there existed input that could trigger \semb. \tool is designed to overcome such limitations of tests. 
\item ($19.6\%$) \textbf{Refactoring.} By \cite{refactoring}, refactoring is the process of restructuring the existing factoring without changing its external behaviors, which is prevalent in Java by \cite{vassallo2019large}. 
In general, the relevant API-level refactoring can be categorized as inter-method and intra-method. Inter-method refactoring, e.g. Extract Method, and Inline Method, is the most common type because it is frequently used to extend API flexibility based on Java polymorphism \cite{polymorph}.

\item ($2.4\%$) \textbf{\semb not triggerable by old input.} The inconsistent behaviors cannot be triggered by the old variables, but only by the variables in the new version. This situation mostly occurs for new functionality, because the new input serves as the option to trigger additional behaviors. 
\item ($14.6\%$) \textbf{Internal \semb has no impact on API output} The changed output of breaking changes has no impact on the output of the API to be observed by users. For example, if the logs are changed, while the return value remains the same, the output of the API is considered unchanged. 
\item ($36.0\%$) \textbf{Benign changes} As allowed by SemVer, benign changes, such as bug fixes, and new functionality, do not substantially break the original semantics. They usually would not break downstream projects, and thus are considered false positives. 
\end{itemize}

We found that except for the first reason (limitation of unit tests), the rest is the false alarm of \semb to be ruled out. For the refactoring, \tool extracts inter-procedural semantic graphs to eliminate syntactic refactoring. For the third and fourth reasons, \tool excludes them by checking the impact of the captured \semb changes. As for the benign changes, we further conduct a study to identify them by patterns.

\subsection{Study of Benign Changes}
\label{study:harmless}

\tool aims at identifying the benign changes to avoid reporting them as breaking. Since the benign changes stipulated by SemVer are not well defined, we summarise the syntactic patterns of benign changes by manually studying commits with such intentions. Then, these patterns will be categorized by their semantic representations on PDG and CFG to be automatically identified. According to \cite{levin2017boosting}, the intentions of commits can be categorized into (1) Bug fixing; (2) Performance improvement; (3) Feature introduction, deletion, and modification.
Since the performance improvement does not explicitly change the output, it would usually not be caught by \tool, thus unnecessary to identify it. As feature deletion and modification are not allowed by SemVer, they should be directly reported instead of being ruled out. Hence, we focused on bug fixes and new functionality.
We collected $584$ commits from the most starred 25 Java projects on Github to summarise the patterns by searching for the keywords in commits, which are \textit{fix}, \textit{correct}, \textit{improve}, \textit{address}, \textit{tweak}, \textit{clean up}, and \textit{add/new feature/functionality}. \ly{Then, the diffs were manually categorized into several patterns.}

\subsubsection{\textbf{Bug Fixes ($393$ cases) Categories}}
\begin{itemize}[leftmargin=9pt]
    \item \textbf{Additional/ conditions and branches.} ($41.85\%$) The change introduces additional conditions to narrow down or broaden the input range. The conditions mostly introduce new branches of statements for additional handling. This is usually to enforce the original rules by disallowing illegitimate input.
    \item \textbf{Changed/deleted conditions and branches.} ($13.28\%$) The original conditions are changed or deleted to fix unreasonable behaviors. Usually, the change handles the illegitimate input to enforce the original rules, but it sometime introduces regression errors.
    \item \textbf{Similar substitution.} ($11.60\%$) Variable types or methods are substituted with similar ones with semantically equivalent method names for better implementation. The original functionality is meant to be maintained.
    \item \textbf{Additional \textit{try/catches}.} ($10.08\%$) It introduces additional \textit{try/catch} pairs or adds \textit{catches} to existing \textit{try}. This kind of change handles more exceptions to avoid unexpected crashes.
    \item \textbf{Assignment revision.} ($10.76\%$) The output assignments are changed partially or completely to correct wrong behaviors or improve sub-optimal behaviors. Such changes can also be found in breaking issues because they could break downstream projects unexpectedly if they are used incorrectly. 
    \item \textbf{Auxiliary variables.} ($6.68\%$) Auxiliary variables are introduced to control the process, which often participates in \textit{if} conditions. For instance, a counter is used to avoid infinite loops.
    \item \textbf{Other.} ($5.75\%$) It consists of \textit{changing internal type, initializing variables, improving method/variable modifiers}, etc. They slightly change syntax without modifying the original semantics. 
\end{itemize}

\subsubsection{\textbf{New Functionality ($191$ cases) Categories}}
\begin{itemize}[leftmargin=9pt]
    \item \textbf{New classes/methods.} ($61.29\%$) New classes/methods are introduced as the entries for new implementations which can be called by the existing APIs.
    \item \textbf{Additional branches.} ($22.58\%$) New implementations/handlers are optionally accessible via new \textit{if} or \textit{switch} branches. 
    \item \textbf{Additional parameters.} ($9.68\%$) New parameters are added to existing internal methods to control or optimize existing functionalities by providing more options.
    \item \textbf{Additional fields.} ($6.45\%$) New fields of returned objects are added with corresponding data and control dependencies. They are used to provide additional information loaded in the fields. 
\end{itemize}

\ly{The new functionality cases were only considered \textit{new} if the original functionalities were intact. Thus, the New Functionality cases were mostly additive changes.} These patterns will be summarized to heuristics based on PDG and CFG. The commit IDs are provided on our website \cite{dataset}.

\section{Methodology}\label{sec:methodology}
According to Gunter et al. \cite{gunter1992semantics}, since the change of semantics can be interpreted to infer the change of output which usually leads to the breaking issues, \tool is designed to infer the \semb by measuring semantic diff.

\begin{figure*}[!t]
  \includegraphics[width=1\linewidth]{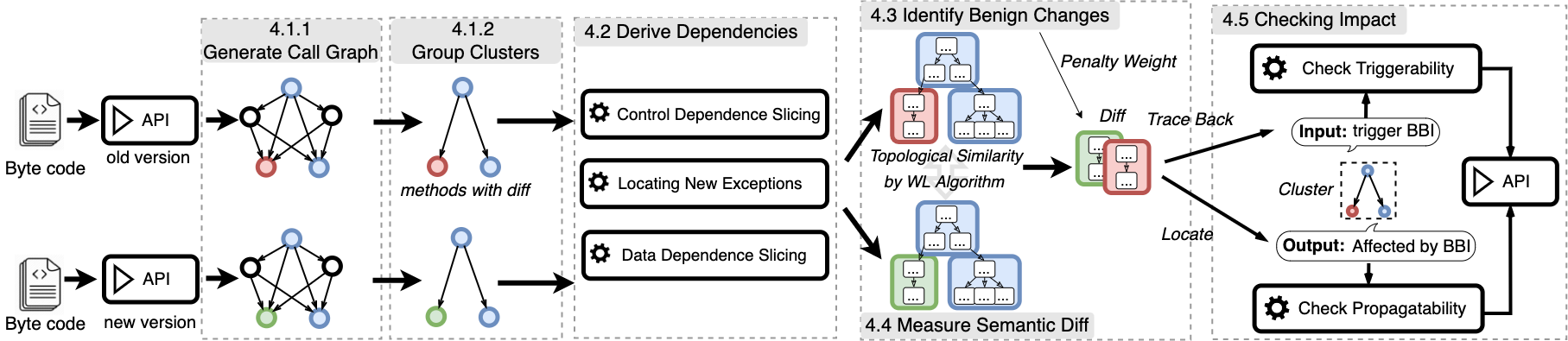}
  \caption{Overview of \tool}
  \label{fig:overview}
\end{figure*}

\tool aims at detecting \semb issues over \pnm upgrades based on APIs. As shown in Figure \ref{fig:overview}, \tool consists of five major steps, (1) \textbf{Group \subgraphs from call graphs}: Given the API pairs, \tool constructs the call graphs of them, from which it groups the consecutive changed methods as \subgraphs. (2) \textbf{Derive Dependencies Summaries}: \tool backward slices the output of \subgraphs to obtain inter-procedural dependency summaries of data/control/exception.
(3) \textbf{Match patterns for benign changes}: we heuristically summarized the patterns from Section \ref{study:harmless} for \tool to identify the statements of benign changes.
(4) \textbf{Measure semantic diff}: Inter-procedural semantic graph pairs are constructed from dependencies summaries. Semantic diff is measured by topological similarity based on subgraph isomorphism by Weisfeiler-Lehman (WL) algorithm. If changes in semantics are too large, the \subgraph is considered to be affected by \semb.
(5) \textbf{Check the impact of \semb}:  
\tool checks if the \semb of each \subgraph is triggerable as well as if the breaking change can propagate back to API's output.

\subsection{Grouping Clusters from Call Graphs}
Given the byte code, \tool first identifies the APIs that are worth checking by narrowing down the API candidates to those with identical signatures only.
Then, it constructs the call graphs for the APIs to derive the subsequent method calls for further analysis.

\subsubsection{\textbf{Generating Call Graphs for Candidate APIs}}
\label{callgraph}
{\tool starts with identifying the set of API pairs that have no \synb. Specifically,  ${A_{cand}} =  \{  \langle api_{old},api_{new} \rangle  \mid Sig_{api_{old}}= Sig_{api_{new}\}}$} is the API pair candidates.
Note that the return type of $Sig_{api}$ has to remain consistent, except it is a widening cast. For example, if the return type of the old API is changed from $long$ to $int$ in the new API, the compilation would succeed after the upgrade. 
To obtain $A_{cand}$, Soot \cite{vallee2010soot} was leveraged to collect API sets $A_{old}$ and $A_{new}$ from class files of the old and the new libraries respectively. 

Based on $A_{cand}$, \tool generates call graphs of each API with Soot by the Spark algorithm \cite{spark}.
For method bodies, Soot transforms them into Jimple \cite{jimple}, a typed IR. The size of the call graph can grow exponentially based on the depth, which results in the inefficiency of the analysis of methods at deep levels.
According to \cite{schroter2010stack}, the semantics of a method decays along the calling chain to be negligible at the depth of around 10 stacks. Thus, conservatively, we set the depth of the call graph $x \in \left [  1, 15\right ]$ to boost the performance.

\subsubsection{\textbf{Grouping Clusters}}
As discussed in Section \ref{noise}, \subgraphs are constructed {by grouping methods that have signature or body changes} to mitigate the inter-method refactoring. Given a pair of call graphs $CG_{old}$ and $CG_{new}$, in terms of the method's signature and body code, methods from both call graphs can be classified as changed methods $M_{changed}$ and unchanged methods $M_{unchanged}$. According to \cite{vassallo2019large}, inter-method refactoring involves methods that are directly connected in a call graph. For instance, ``Extract method'' creates a new method to replace the removed block. Thus, \tool groups as many directly connected (consecutive) methods as possible of $M_{changed}$ as \subgraphs. The \subgraphs are analyzed and sliced altogether as a whole to neutralize the inter-method refactoring.

\subsection{Deriving Dependencies Summaries}
\label{sec:detect}

For a \subgraph pair  $\left \langle c,c' \right \rangle$, \tool relies on three kinds of inter-procedural semantic summaries to model the relationship between the input and output, which are Data Dependency Summaries (DDS), Control Dependency Summaries (CDS), and Exception Summaries (ES). Each dependency in the summary is embodied as a Static Single Assignment (SSA) statement from the IR. These summaries are used to capture the factors that potentially change the output regarding the data calculation, control logic, and exception handling. Listing \ref{lst:exp1} will be used as a running example.

The DDS and CDS of the output of \subgraphs are extracted based on backward slicing of the output in the Program Dependence Graph (PDG) recursively. We first define \subgraph $c = \{m_{root}, m_j\mid j=0,...,n\}$ where $m$ is the method, and $n$ is the number of methods, excluding the root. The ES is a set summarised from the unhandled exception exits of all methods within a \subgraph.
Next, we define the output of a \subgraph as the non-local variables that are written after the execution of the \subgraph. The output can be in three forms: (1) Variables returned by $m_{root}$; (2) Class fields written in case $m_{root}$ returns void; (3) Exceptions thrown, as exception variable is a special non-local variable to be handled out of $c$. In general, the output is considered as the impact that the $c$ imposes on the global environment.  
The semantic dependency summaries are robust against syntactic intra-method refactoring, such as moving, renaming, pushing down/pulling up, and splitting/merging local variables, because \tool directly models the relationship among non-local variables without the interference of local variables.

\subsubsection{\textbf{Data Dependencies Summary}}
\label{dds}
DDS is used to model the relationship between the input and output as an aspect of data calculation. Based on PDG, output statements are backward sliced to derive data dependence statements.
If any non-local variable, such as parameter, is met, \tool associates all statements met before with the non-local variable to check the triggerability later. Since the operands in SSA are prone to renaming, the operands are normalized as \textit{var}, \textit{parameter}, and \textit{field} based on their roles.

To support inter-procedural analysis, when \tool meets a called method $m_0$ within the \subgraph during the slicing, it dives into $m_0$. In $m_0$, the operations saving is executed with the same pattern as the root method $m_{root}$. The only exception is that parameters of $m_0$ are mapped to the local variables in $m_{root}$ instead of non-local variables. 
If \tool meets a method not included in the \subgraph, which are either unchanged methods or methods from other libraries, such as Java built-in classes, \tool does not dive into it, as analysis of them does not make difference on \semb detection.

\begin{figure}[!t]
  \includegraphics[width=1\linewidth]{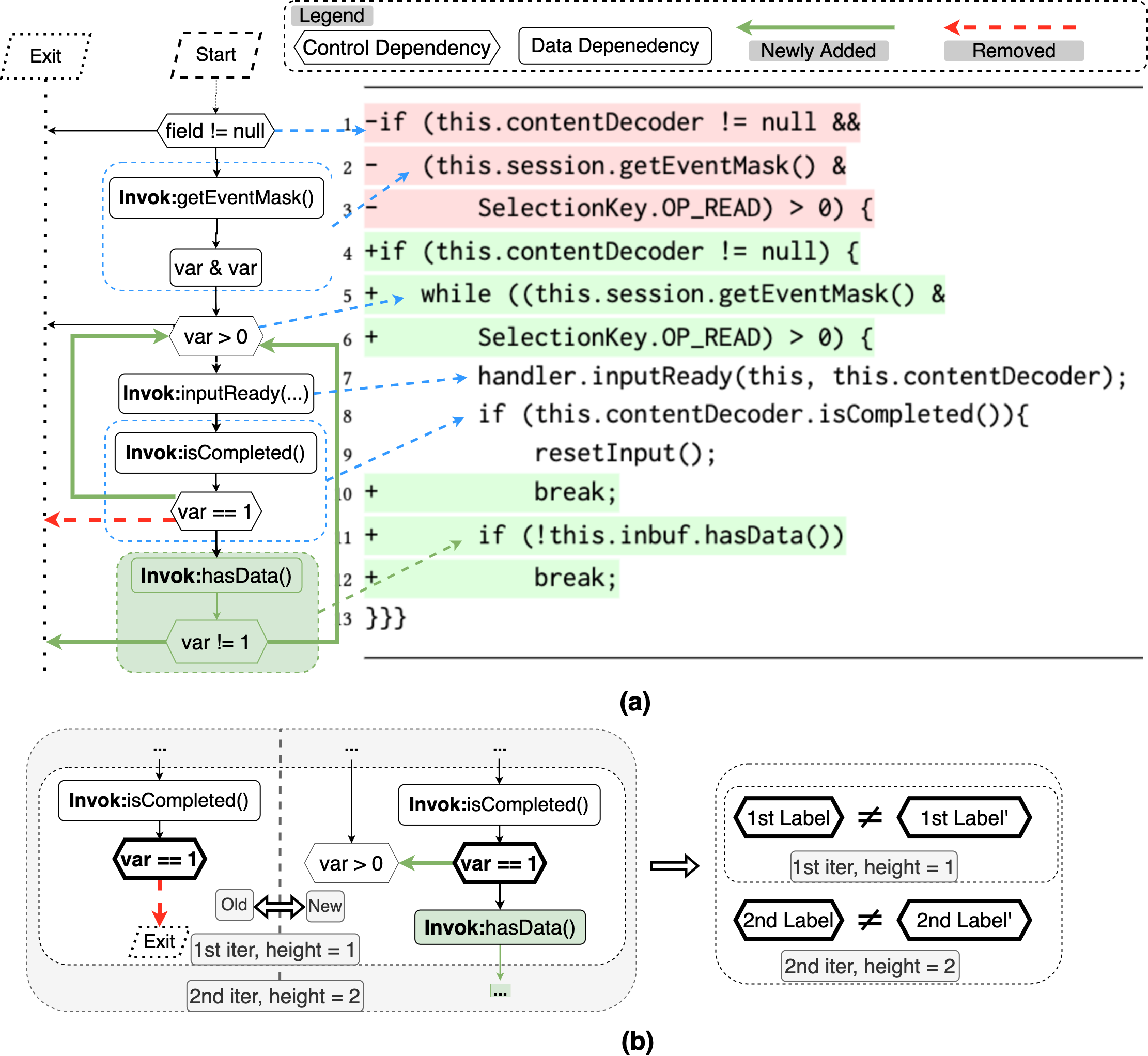}
  \caption{(a) Data \& Control Dependencies Summaries in the semantic graph of Motivating Example (b) Example of subgraph comparison of Node \textit{var==1} at height 1 and 2} 
  \label{fig:dds}
\end{figure}

\subsubsection{\textbf{Control Dependency Summary}}
\label{sec:cds}
CDS describes the controlling semantics of the \subgraph, which determines the execution branching. Towards a pair $\left \langle c,c' \right \rangle$, CDS is derived by backward slicing the output of $m_{root}$ with Control Dependency edges recursively. Boolean conditions, e.g. \textit{if}, directly and indirectly, lead to the output will be collected. The inter-procedural analysis is conducted in the same manner as DDS. 
To normalize the conditions against syntactic changes, e.g. refactoring, \tool calculates the Disjunctive Boolean Summary (DBS) which is a joint logic symbol of the original logic and the reversed one. For instance, $a>0$'s DBS is $\langle >0 \mid <=0 \rangle$. Moreover, the DDS of variables used in the conditions is extracted to capture the possible changes in variable values. These Data dependency statements are associated with the conditions.

Recall the motivating example, the DDS and CDS are extracted based on the output (L7 \textit{handler}) slicing in Figure \ref{fig:dds}. These summaries on the left are organized based on the control flows from CFG. Since only L11-L12 introduces new conditions, the CDS changes, but DDS does not change. The switch from \textit{if} (L2) to \textit{while} (L5) changes control flow without changing CDS or DDS. Hence, It is necessary to include CFG in the analysis to capture all possible semantic changes. 

\subsubsection{\textbf{Exception Summary}}
According to the study from \cite{mostafa2017experience}, unexpected exception throwing is a common \semb issue, which is caused by un-handled new exceptions. Thus, it is necessary to check if the exceptions are consistent during the upgrade for the \subgraph. Because exceptions have no data dependencies, but only control dependencies, the exceptions extracted from the \subgraph are associated with the corresponding conditions from the CDS.

\subsection{Matching Patterns for Benign Changes}
\label{harmless}

In order to match benign changes in PDG and CFG. We first summarize semantic patterns from the syntactic source code patterns from Section \ref{study:harmless}. Given a \subgraph $c$, statements of benign changes are identified and collected to a set $S_{ben}^c$. 
According to Campos et al. \cite{campos2017common} and Pan et al. \cite{pan2009toward}, there were 9 categories of bug fixes patterns in Java. We further generalize them into 5, namely, from major to minor, controlling conditions (CC), field, method call, variable assignment, and try/catch. Also, we found the patterns of new functionality rely on the first 3 objects. Since these categories that can be identified from CFG/PDG based on IR could cover most cases of benign changes, heuristics to identify the overall benign changes were designed based on these categories.

\begin{itemize}[leftmargin=9pt]
    \item \textbf{CC Adjustment:} The benign changes adjust the conditions for legitimate use. If the subsequent implementations along the branch are not changed, the change is considered benign. \textbf{Steps}: (1) Get changed conditions from two lists of old and new conditions of CDS from Section \ref{sec:cds}. (2) Locate subsequent blocks of changed conditions by CFG and compare the DDS statements in those blocks to check if changed conditions lead to the same implementation. (4) If so, add the changed conditions as well as the backward sliced data dependency statements of them to $S_{ben}$, because they all contribute to the adjustment.
    \item \textbf{CC and \textit{Try/Catch} Extra Handling:} Because the old implementation is not sufficient to cover all legitimate input, developers may add extra handling to the original to expand the input domain. The extra handling can either fix a corner case or introduce new functionality to support a new option. But in either way, they are reflected as extra branches in exceptional CFG. \textbf{Steps}: (1) Locate only the new conditions in CDS and new \textit{Catch} in ES. (2) Derive statements in two subsequent blocks of boolean conditions and \textit{Catch} by CFG. (3) Since the extra handling would not tamper with the original implementation, one of the blocks should remain identical and the other one is new. Compare the blocks with the old implementation to get the new ones. (4) Add these new conditions/exceptions, statements from new blocks, and their data dependency statements to $S_{ben}$. 
    \item \textbf{Field: Augmented Output:} In Java, the instance as output is augmented to accommodate more fields. Another case is non-static methods returning void set more fields as the output. \textbf{steps}: (1) Identify new fields by comparing the output's old and new field lists by name and type. \ly{Note that \tool only captures the newly added fields, which means the old fields should remain the same. If the number of output fields does not increase in the new version, it is not considered augmented, but possible breaking changes, as the old implementation could rely on the types of original fields.} (2) Backward walk through the def-use chains of the fields recursively to add relevant statements to $S_{ben}$. Backward slicing does not work in this case, because it returns dependencies of the output instance, instead of specific fields. Hence, fine-grained slicing at the field level is applied.
    \item \textbf{Method Call: Similar Substitution:} Classes or methods are substituted with similar analogies to enforce the original rules. In this case, the semantics of the substituted component is reflected by its role in the context. For example, if one statement of invocation is replaced, but the contextual statements remain the same, the role of the substitution is not changed. \tool captures this semantics by using a neighbor-preserving algorithm in Section \ref{semdiff}, which considers two nodes are identical if all $1_{st}$ neighboring nodes remain the same.
    \item \textbf{Variable Assignment Revision:} Some variables are assigned different values or new assignments are included. If the variable is local and not a data dependency of the output, it is likely to be an auxiliary variable that controls correct behaviors without directly tampering with the value of output. It is identified by (1) Obtain data dependency statements of conditions from CDS. (2) Compare them based on each condition to get new statements, and check if they belong to new variables. (3) If so, add them to $S_{ben}$. If the output or data dependency of output is assigned different values, we do not explicitly classify them to $S_{ben}$, because it is likely to introduce breaking issues by yielding abnormal output.
\end{itemize}

Theoretically, benign changes should be identified and ruled out to avoid false positives. However, the identification cannot be perfectly accurate and they sometimes still cause \semb, such as the regression issue, because either developers fail to anticipate the breaking or the downstream projects use the API illegitimately. Thus, \tool fuzzily measures the semantic diff to determine the \semb  with the de-emphasized benign changes instead of completely ignoring them. However, if the accumulated semantics is changed greatly, it is still considered \semb.

\subsection{\textbf{Measuring Semantic Diff}}
\label{semdiff}
To measure the semantic diff for a \subgraph pair, \tool constructs an inter-procedural semantic graph as Figure \ref{fig:dds} by connecting Exception/Data/Control Dependencies Summaries with execution paths from CFG. The semantic graph preserves the execution logic among relevant dependency statements to model the behaviors of non-local variables of \subgraphs. \tool infers the extent of semantic change by calculating the topological similarity of semantic graphs by subgraph matching algorithm based on Weisfeiler-Lehman (WL) graph kernel \cite{shervashidze2011weisfeiler} with weighted statements of $S_{ben}$. If the final value is above a threshold, the \subgraph is affected by \semb.

Here are the reasons for the adoption of the WL graph kernel. WL graph kernel can convert the original graph to a sequence of substructures defined as kernels that sort and compress topological and labeling information of adjacent nodes. The kernel pairs preserving the neighboring semantics can be used to calculate the semantic similarity based on the number of matched kernels. Besides, the runtime scales linearly in the number of edges better than other kernels, such as Random-Walk or Shortest-Path \cite{shervashidze2011weisfeiler}. WL graph kernels are designed for directed discrete large graphs which suit the scenario of semantic graphs.

Inspired by the graph matching algorithm of PDG in CCGraph \cite{zou2020ccgraph}, \tool relies on the subgraph isomorphism based on \textit{h} iterations of WL graph kernel calculation to determine the semantic diff between semantic graphs $\left \langle G,G' \right \rangle$. CCGraph targets detecting code clone only based on PDG, while the semantics from PDG is not sufficient for \semb detection. For example, the crucial control flow change in Listing \ref{lst:exp1} is not reflected in PDG. Thus, \tool measures the semantic diff between sliced statements connected by control flows for better semantic representation. Also, \tool further de-emphasizes the benign changes to align with SemVer rules. 
The procedures are described below in Algorithm \ref{alg:1}:
\begin{itemize}[leftmargin=9pt]
    \item Labels of nodes in graphs are hashed as the initial values. 
    \item In \textit{$i_{th}$} iteration of WL algorithm \cite{shervashidze2011weisfeiler}, labels of each node as well as its neighbors in \textit{i} hops are compressed into a new label by local sensitive hashing according to WL algorithm.
    \item In \textit{$i_{th}$} iteration, if labels of two nodes are identical, the subgraphs of the node pair are considered as isomorphic at height \textit{i}. 
    \item After all iterations, the number of non-identical node pairs multiplies with a deteriorating weight $w$ and benign penalty $p$ per pair as the final graph kernel value $K$. $K$ is normalized to compare against the threshold $T$. If $K>T$, the \subgraph pair has \semb. $K$ is calculated as 

\end{itemize}
\begin{flalign}
    \label{eq:k}
    \scalebox{1}{$K = \frac{\sum^{h}_{i=1} \left \langle l(n)|l(n)\neq l(n') \right \rangle * (h-i+1)/h * sizeof(S_{ben})/sizeof(G')}{min(sizeof(G), sizeof(G'))}$}
    \nonumber
\end{flalign}

Same as \cite{zou2020ccgraph}, deteriorating weight $w$ is calculated as $(h-i+1)/h$ for $i_{th}$ iteration, because the closer the neighbors of node $n$ are, the more they affect the node $n$. The benign change penalty $p$ is calculated as $sizeof(S_{ben})/sizeof(G')$ to dynamically adapt to the size of $G'$. \ly{Since lower $T$ lowers the precision of \tool, while higher $T$ lowers the recall, the $T$ is empirically set as $0.1$ to balance the precision and recall.} In Figure \ref{fig:dds}(b), the node \textit{var==1}'s neighbors are converted to subgraphs and then compressed by WL algorithm to form a label at height $1$. Height $2$ is formed in the same way. Evidently, none of the label pairs of node \textit{var==1}' is identical.

\begin{algorithm2e}[t]
    \footnotesize
 \setcounter{AlgoLine}{0}
 \caption{{Algorithm of Measuring Semantic Diff}}
 \label{alg:1}
 \DontPrintSemicolon
 \SetCommentSty{mycommfont}
 {
     \KwIn{$\left \langle c,c' \right \rangle$: \subgraphs pair with nodes $n$, label $l(n)$. $h$: iteration number of WL algorithm. $T$: threshold, $w_i$: deteriorating weight at $i_{th}$, $p$: benign change penalty. }
     \KwOut{$R$: Result of existence of \semb in $\left \langle c,c' \right \rangle$}
     \ForEach{$i_{th}$ iteration in $h$}{
        \ForEach{node $n$ in \subgraph $c$}{
            $sg(n) \gets WLGenSubGraph(n, neighbor(n, i_{th}))$\;
            $L(n) \gets \sum{(l(n)|n\in sg(n))}$\;
            $L(n) \gets sort(L(n))$\;
            $l(n) \gets l(n)+L(n)$\;
            $l(n) \gets WLcompress(l(n))$\;
        }
        $w_i \gets (h-i+1)/h$\;
        $k_i \gets sizeof(l(n)\neq l(n'))*w_i*p$\;
    }
    $k \gets \sum{k_i/min(sizeOf(c,c'))}$\;
    \If{$k>T$}{
        $R \gets 1$\;
    }
    \KwRet{$R$}   
}
\end{algorithm2e}

\subsection{Checking Impact of Semantic Breaking}
\label{trigProp}
This procedure only proceeds when a \semb \subgraph is caught in the previous step. Only if both triggerability and propagatability are feasible, the \semb \subgraph is considered a threat to the API.

\begin{figure}[!t]
  \includegraphics[width=0.9\linewidth]{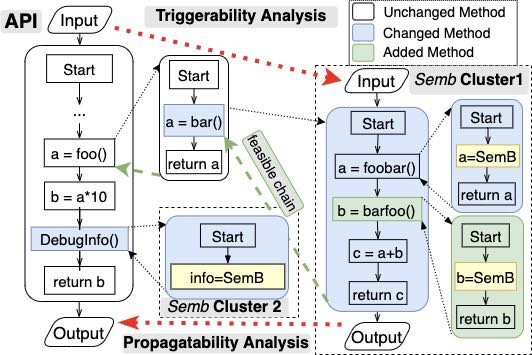}
  \caption{Triggerabitliy and Propagatability Analysis}
  \label{fig:trigprop}
\end{figure}

\subsubsection{\textbf{Verifying Triggerability}}
\label{trigger}
Triggerability defines if \semb can be triggered by old input. In Section \ref{sec:detect}, during the backward slicing, input is associated with relevant statements. If the statements are caught as the \semb changes, the associated input would be checked, including parameters and fields. If the problematic input is introduced in the new version, the \semb cannot be triggered, because the old implementation cannot access the new input.

\subsubsection{\textbf{Verifying Propagatability}}
\label{prop}
Propagatability determines if the \semb output can propagate from the \subgraph $c = \{m_{root}, m_j\mid j=0,...,n\}$ along call paths to API to affect users. Since \semb is introduced in the new version, \tool only verifies the propagatability in the call graph of the new version. \tool uses JGraphT \cite{michail2020jgrapht} serving as the graph infrastructure which first derives all call paths by Dijkstra Algorithm \cite{dijkstra} from the API to the $m_{root}$ as $P = \{path_{i}\mid i=0,...,n\}$. Second, For each call path, \tool calculates PDG for every method along the call path to verify if the inter-procedural dependency chain is feasible along every PDG from $m_{root}$ to API's entry method. Finally, propagatability $Pg$ is obtained by $iff \exists path\in P, depChain (path)==True, Pg=True$. 

For example, in Figure \ref{fig:trigprop}, \textit{\subgraph1}'s output is propagatable to the API's output. The output of $m_{root}$ of \textit{\subgraph1}, $return\, c$ has a feasible dependency chain $a=bar()\rightarrow a=foo() \rightarrow return\, b$ as indicated by the green dotted line. Thus, the output of the API indirectly depends on the \semb change in \textit{\subgraph1}. In contrast, an example of an un-propagatable \semb change is \textit{\subgraph2} which yields debugging information not used as the API's output.

\section{Evaluation}\label{evaluation}

\tool was implemented in 9.2K LOC based on Jimple IR of Soot 4.2.1 in Java. 
We aim at answering  the following research questions:

\noindent
\textbf{RQ1}: What is the accuracy of \tool in terms of \semb detection?

\noindent
\textbf{RQ2}: How is the effectiveness of \tool against unit tests?

\noindent
\textbf{RQ3}: How do top Java libraries comply with SemVer rules?

\subsection{Evaluation Setup}
\label{data}
\subsubsection{\textbf{For RQ1.}} To evaluate the accuracy of \tool, we first construct a high-quality ground truth dataset for the benchmark with other API checking tools and the baseline.

\noindent\textbf{Benchmark Dataset Collection.} Since \tool is designed to detect \semb across \pnm upgrades, we collected broken API pairs of \pnm upgrades from the 20 most used Maven libraries. The steps are (1) We located Github repositories of those libraries. (2) We conducted regression tests by running unit tests from the old versions against the implementations in the new versions to detect \semb. (3) After ruling out the \synb, the failures that caused \textit{AssertionError}, unexpected exceptions, and crashes are considered as \semb, because \textit{AssertionError} means the program is executed abnormally. Since Mostafa et
al. \cite{mostafa2017experience} conducted the same regression tests on some of  the libraries before 2018, we extended the dataset by extracting APIs from their testing logs in the same steps. Eventually, we derived 308 API pairs with \semb from 77 version pairs. For the dataset of compatible API pairs aligning with SemVer rules, we used the dataset from Section \ref{noise}. $363$ API pairs with binary changes without \semb serve as the negative data set of \semb detection. 

\noindent\textbf{Metrics.} The outcomes of \tool are categorized into (1) True Positive (TP): APIs reported have \semb. (2) False Positive (FP): APIs reported do not have \semb. (3) True Negative (TN): The API not reported by \tool has no \semb. (4) False Negative (FN): The API not reported by \tool actually has \semb. Precision, recall, and F-measure are used as evaluation metrics.

\subsubsection{\textbf{For RQ2:}}
We conducted another experiment to verify the effectiveness of detecting \semb APIs over version pairs between \tool and the commonly used \semb detection solution, unit tests.
As the number of APIs of popular libraries is considerable ($406,826$ APIs of 77 pairs), the efforts of manual ground truth checking would be overwhelming. Hence, we selected the top 4 most used libraries with 1 \semb \textit{Patch} version pair each, as they are more likely to have the best testing coverage. In total, $3,846$ APIs were collected. 

\subsubsection{\textbf{For RQ3:}}
All semantically successive version pairs published in the last 20 years of 21 most used Java TPLs from the Maven repository \cite{mvnrepo} were collected for the large-scale analysis. In total, $546$ version pairs and $1,629,589$ APIs were tested. To analyze the compliance of the SemVer rules, we classified them into (1) \textit{Patch}: 334 pairs; (2) \textit{Minor}: 163 pairs; (3) \textit{Major}: 49 pairs. The versions collected were stable unless none was available.

\subsection{RQ1: \semb Detection Accuracy}
\label{rq1}
To evaluate the accuracy of \tool against existing tools, we have selected 5 Java API compatibility checking tools (i.e., revapi \cite{revapi}, japicc \cite{japi-compliance-checker}, japi-checker \cite{jchecker}, clirr \cite{clirr}, sigTest, \cite{sigtest}), which are commonly used in benchmarks \cite{jezek2017api} and industrial software, such as Apache Httpclient~\cite{httpclient}. Besides existing tools, we also implemented a baseline tool that relies on the same call graph construction procedures as \tool, but \semb is considered as positive if any binary diff exists in any method called by APIs.

Table \ref{tab:ibi} presents the accuracy evaluation of \tool and other tools on the benchmark data set from Section ~\ref{data}. \tool achieves $90.26\%$ recall, $81.29\%$ precision, and $85.54\%$ F-measure. It is concluded that \tool outperformed other tools because these tools could only detect \synb instead of the \semb.
There were two reasons why some tools could still have TP. First, tools, such as revapi, not only evaluated the compatibility based on signatures but took other features, such as method accessibility and abstraction into account. 
Second, some of the tools only returned the incompatible class names without method names.
If any API signature from the data set had the same class name as the returned class names, we considered the \semb API was detected.

\noindent\textbf{False Negative Cases Discussion.} We manually examined the 30 false-negative cases and summarised 4 reasons why \tool made false decisions. A detailed name list is provided in \cite{dataset}.

\begin{table}
  \caption{Benchmark Accuracy of \semb Detection based on APIs}
  \label{tab:ibi}
\scalebox{0.95}{
\begin{tabular}{lcccccc}
    \toprule
\textbf{Tools}   & \textbf{TP} &\textbf{FN}  & \textbf{Recall} & \textbf{FP} & \textbf{Precision} & \textbf{F-measure}                    \\
    \midrule
\textbf{\tool}     & \textbf{278} & \textbf{30} & \textbf{90.26\%} & \textbf{64} & \textbf{81.29\%} & \textbf{85.54\%} \\
\textit{baseline}     & 302 & 6 & 98.05\% & 363 & 45.41\% & 62.07\% \\ 
revapi       & 30 & 278 & 9.74\% & 21  & 58.82\% & 16.71\% \\
japicc       & 21 &287 & 6.82\%  & 14  & 60.00\% & 12.25\% \\
japi-cker & 14 &294 & 4.55\%  & 10  & 58.33\% & 8.44\% \\
clirr        & 1 & 307  & 0.32\%  & 0  & 100.00\% & 0.64\%  \\
sigTest      & 0 & 308  & 0.00\%  & 0   & N.A.  & N.A. \\
    \bottomrule
\end{tabular}}
\end{table}

\begin{itemize}[leftmargin=9pt]
    \item ($46.67\%$, 14 cases) \textbf{Secondary Output}: It is the output embodied as debugging messages, written files, and other auxiliary information conveyed by the APIs. Unlike the primary output, the secondary outputs are not global variables passed to the users' programs, but the unit tests sometimes still include them by \textit{assertion}. Since, unlike the secondary output, users directly use the primary one which could break downstream projects, \tool focuses on the primary output to cover the mainstream scenarios. 
    \item ($23.33\%$, 7 cases) \textbf{Falsely Identified Benign Changes}: The patterns of these changes fall into the summarised benign changes. However, no evidence suggested they are benign from commit messages or documentation. Although de-emphasizing benign changes is not perfect, it does improve the precision at the relatively low cost of false negatives.
    \item ($20.00\%$, 6 cases) \textbf{Signature Reflection}: The breaking was caused by different signature reflections \cite{reflection}. \tool cannot capture the behaviors that can only be triggered dynamically, and thus can complement testing for a more comprehensive detection.
    \item ($10.00\%$, 3 cases) \textbf{Subtle Changes}: The changes were too subtle to be detected by the semantic diff measuring, such as the change of index number of an array, which may not always cause changed output. Although the threshold may overlook some subtle breaking changes, it reduces many more false positives.
\end{itemize}

\noindent\textbf{False Positive Cases Discussion.}
$64$ false-positive cases were listed with 3 reasons why the results were incorrect.
\begin{itemize}[leftmargin=9pt]
    \item ($53.13\%$, 34 cases) \textbf{Large Accumulated Semantic Diff}: Although \tool already assigned low weights to the benign changes, multiple benign changes can still have a considerable stacked impact on measuring semantic diff. Because benign changes cannot be accurately identified and filtered out, we have to trade off between the under-fitting and over-fitting by weighting.
    \item ($37.50\%$, 24 cases) \textbf{Equivalent Re-implementation}: The code with the same functioning was re-implemented in another way in the new version. One example is the Java feature evolution. Java 8 \cite{java8} introduced a feature, \textit{stream}, for parallel aggregate operations. Although it can be used with \textit{forEach} to work the similar way as primitive \textit{for loop}, they are written in totally different byte code. Another example is the change from \textit{array.getElementAt(n)} to \textit{array.subString(n, n+1).get(0)}. Both of them return the same element in the array, but they are implemented in different ways, which resembles the \textit{Type 4 Code Clone} problem that hardly has efficient and effective solutions so far. Hence, \tool fails to identify the semantic equivalence between them at the current abstraction level.
    \item ($9.38\%$, 6 cases) \textbf{Unhandled Exception}: They were detected as BBI APIs because the newly thrown exceptions are not correctly handled in the new version. But they are actually handled by the super-type exception catchers. \tool checks the exception handling by comparing the exception signatures of the thrown and the catcher. If they are not the same or the catcher is not a general exception, such as "Exception", the thrown exception is considered not caught. In fact, if the thrown is an inherited sub-type of the catcher with different signatures, the thrown can still be caught. \tool made such wrong decisions due to the lack of knowledge of the exception inheritance hierarchy.    
\end{itemize}

The \textbf{baseline} tool achieved high recall but low precision, which failed to detect 6 cases of signature reflection due to its static basis. All APIs from the compatible test set were false positives.

\vspace{1mm}

\newenvironment{boxedtest}
{\noindent\begin{Sbox}\begin{minipage}{\linewidth-7.5\fboxrule-2\fboxsep-1pt}}
{\end{minipage}\end{Sbox}\doublebox{\TheSbox}}
\begin{boxedtest}
\textbf{Conclusion of RQ1: }
\tool outperformed other API compatibility checking tools and achieved $90.26\%$ recall, $81.29\%$ precision, and $85.54\%$ F-measure in terms of \semb API detection. \tool achieved much better precision than the baseline tool ($45.41\%$), which indicates that \tool is able to effectively filter out the false-positive changes.

\end{boxedtest}
\subsection{RQ2: Effectiveness of Detecting \semb against Unit Tests}
As unit tests are widely used to detect \semb based on APIs, we compared \tool against unit tests regarding the number of detected \semb APIs. A similar tool, DeBBI, was proposed by Chen et al. \cite{chen2020taming} to detect \semb based on augmented unit tests, but it requires manual analysis, and no public data or source code is available. Hence, DeBBI is not involved in the evaluation.
Based on the selected libraries from Section \ref{data}, we first obtained APIs that have binary change as set $A_{changed}$. Then, we derived all tested methods from the testing source code. Because developers would not explicitly mention what methods or classes are tested, we made an overestimation by assuming all public and instantiable classes used in the tests along with their public methods are tested. Based on this assumption, we processed the testing classes in the following manner: (1) Irrelevant testing methods were filtered out according to the rules of testing frameworks. For example, if a framework, Junit, was used, methods annotated with $@before$, $@after$, $@ignore$ would be ignored. 2) From the relevant testing methods, we constructed call graphs for each of them, then directly called methods that meet the aforementioned conditions were collected. (3) During running the tests, the dynamic call graphs were calculated to derive the dynamically called APIs. APIs collected are formed as a set $A_{tested}$ to denote the changed APIs covered by tests. 

\begin{figure}[!t]
  \includegraphics[width=1\linewidth]{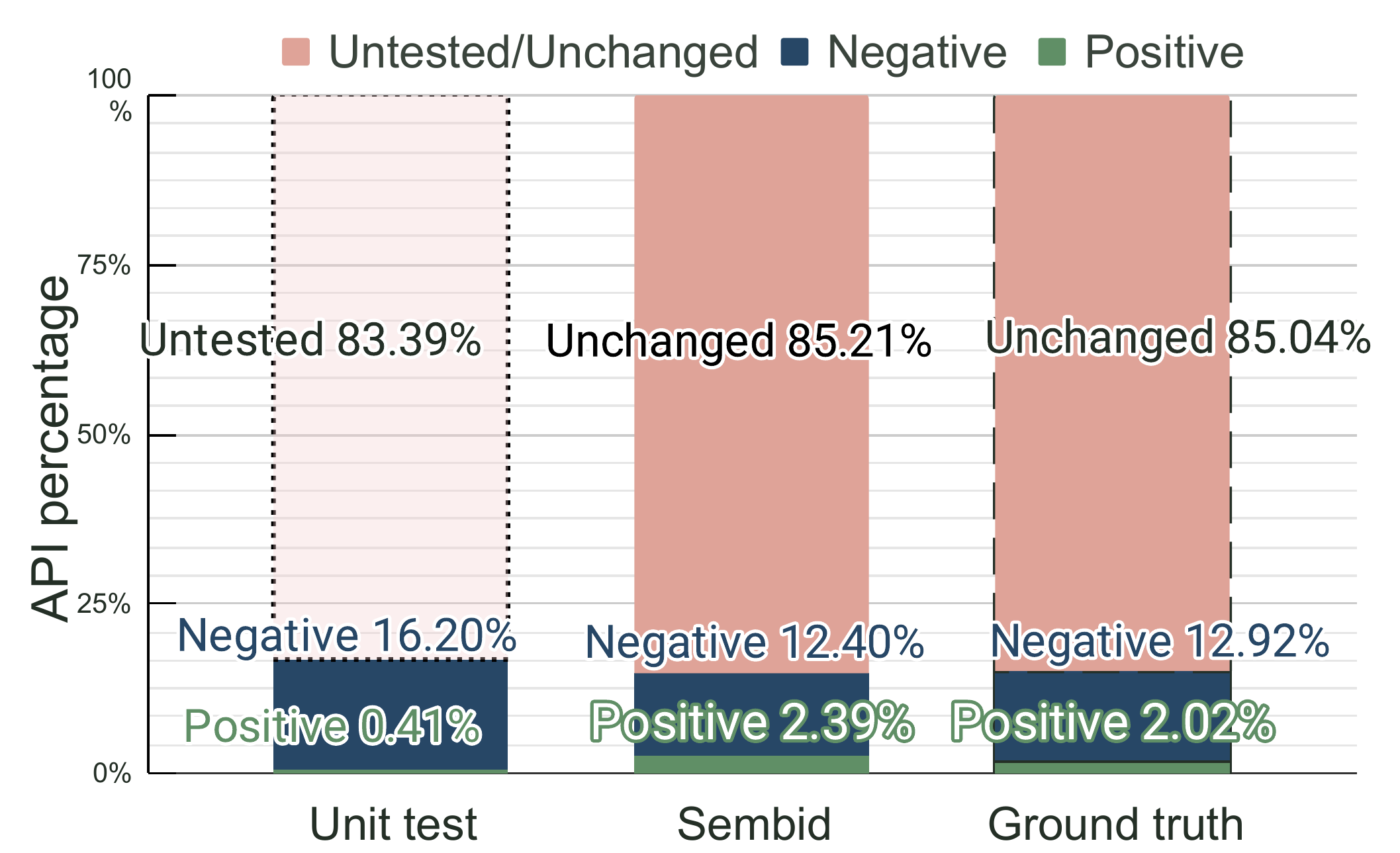}
  \caption{Comparison between unit tests and \tool over all APIs during upgrades as well as the ground truth}
  \label{fig:rq4pie}
\end{figure}

In Figure \ref{fig:rq4pie}, the results are illustrated. In total, $3,846$ APIs were evaluated in 4 version pairs. The ground truth of them was manually confirmed. It is evident that from the first bar unit tests covered averagely $16.61\%$ of APIs, while the second bar indicates that \tool can cover $100\%$ APIs and focus on detecting potential \semb in $15.00\%$ ($577$) APIs with binary changes. The unchanged APIs are naturally compatible. Apart from the coverage of APIs, even for APIs covered by unit tests, unit tests only uncovered $0.48\%$ ($16$) \semb APIs of all APIs. According to the ground truth denoted by the third bar, unit tests failed to detect $79.46\%$ ($62/78$) of \semb APIs, but \tool successfully covered $92.30\%$ ($72/78$) of \semb APIs. Although \tool made some false alarms ($21.73\%=20/92$), generally \tool detected more \semb APIs than the unit tests with 4.5 times more TP. 
However, \tool as a static tool has its limit. As unit tests can dynamically detect \semb in certain APIs with reflection, but \tool is not able to cover them statically.
\\
\\
\begin{boxedtest}
\textbf{Conclusion of RQ2: }
Over 4 version pairs, $3,846$ APIs in total, \tool detected 4.5 times more \semb APIs than unit tests ($72$ v.s. $16$) and achieved better coverage of APIs ($100\%$ v.s. $16.61\%$). Unit tests are able to detect $6$ more \semb caused by dynamic operations than \tool. It indicates that \tool can serve as the complement of unit tests to detect more \semb for \pnm upgrades.
\end{boxedtest}

\subsection{RQ3: Study of Compliance with SemVer}
\label{rq4}

To verify the compliance of SemVer rules in popular Java TPLs, we evaluated the TPLs at library, version pair, and API levels respectively. During the evaluation, both \synb and \semb are considered as evidence of breaking (either \semb or \synb is a breaking). For each version pair, the APIs affected by \synb and \semb as well as the APIs with binary changes were collected. The detection of \synb depends on the aforementioned API checking tools. The \synb APIs are the union of them. The \semb was detected by \tool. The changed APIs were collected with Soot and BCEL.

\begin{figure}[!t]
  \includegraphics[width=1\linewidth]{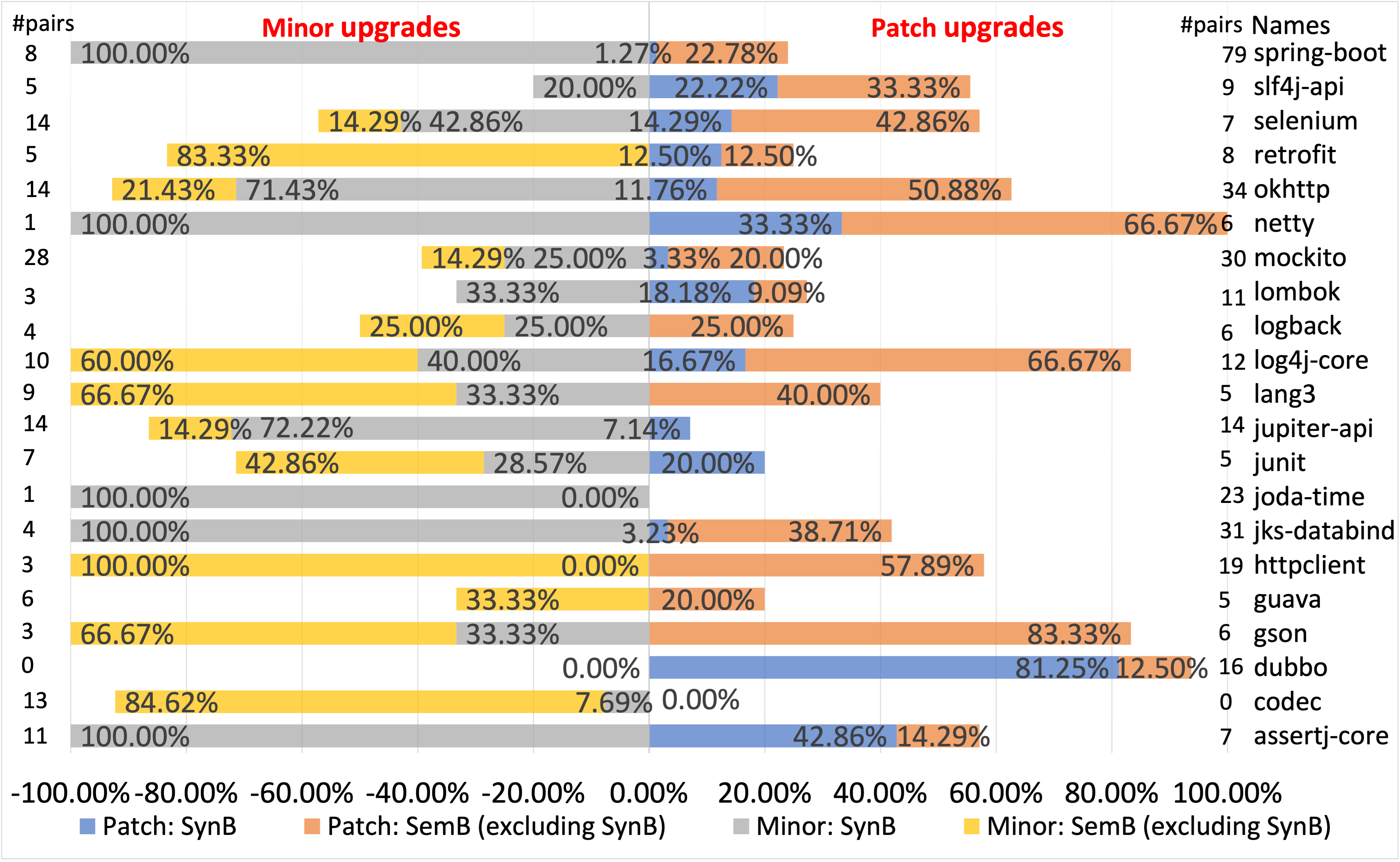}
  \caption{Proportions of version pairs affected by \semb and \synb of \pnm upgrades}
  \label{fig:rq4api}
\end{figure}
Beginning with library and version pair levels, Figure \ref{fig:rq4api} illustrated the proportion of \semb and \synb version pairs of 21 libraries. If one breaking API was detected in a version pair, this version pair was considered to be broken. Note that the \semb version pairs were counted when the version pairs were free from \synb. In other words, the sum of \semb proportion and \synb proportion is the proportion of all broken version pairs. The left side is the \textit{Minor} upgrades, and the right side is the \textit{Patch} upgrades. The numbers of version pairs are annotated at the ends of bars. We found that \begin{itemize}[leftmargin=9pt]
    \item \textbf{Patch upgrades on average}: $14/21 (66.67\%)$ libraries were subject to \synb for at least one version pair ($1.27\%-81.25\%$). $19/21$ $(90.48\%)$ libraries have at least one breaking version pair. It is seen that SemVer rules are hardly applied to popular libraries. However, the situation is getting better at the version pair level. On average, \synb affects $10.45\%$ of version pairs, but \semb affects additional $23.35\%$, which makes $33.83\%$ of version pairs affected by either breaking. The \semb version pairs are over 2 times of \synb. It suggests that \semb detection is necessary in \pa upgrades.
    \item \textbf{Minor upgrades on average}: All $20$ libraries with minor upgrades are affected by any breaking. It is observed that developers are more likely to include breaking changes in \mi upgrades than \pa. At the version pair level, \synb affects $37.42\%$ of version pairs, and \semb additionally affects $26.99\%$, which makes $64.42\%$ of either breaking. Due to legacy reasons, libraries adopt various version release strategies, and many libraries' administrators allow breaking changes in \mi upgrades. 
    \item \textbf{Particular libraries}: The performance varies tremendously among libraries, because they adopt different, even opposite version release strategies. Some libraries have a high ratio of breaking version pairs, such as \textit{dubbo}, \textit{netty}, \textit{log4j}, \textit{gson}. Because they have very few or $0$ \textit{Minor}/\textit{Major} upgrades, they frequently make \textit{Patch} upgrades. They usually make \ma upgrades cautiously when the entire structure is rewritten, such that \textit{log4j} upgrades from 1.x to 2.x, \textit{httpclient} 4.x-5.x. Then almost all normal upgrades with breaking changes were accommodated in \pnm upgrades. If \mi upgrades are rare, \pa upgrades would be mostly broken. For this kind of library, SemVer rules are not properly applied so users have to identify the strategies case by case. There are also some libraries adopting the opposed strategy. For instance, \textit{guava} made \ma upgrades frequently ($13$ times in 7 years) with a few \pa upgrades. Although \textit{guava} still has a few unexpected \semb, SemVer rules are basically well applied, thus users can upgrade it by SemVer rules with ease.   
\end{itemize}

\begin{figure}[!t]
  \includegraphics[width=1\linewidth]{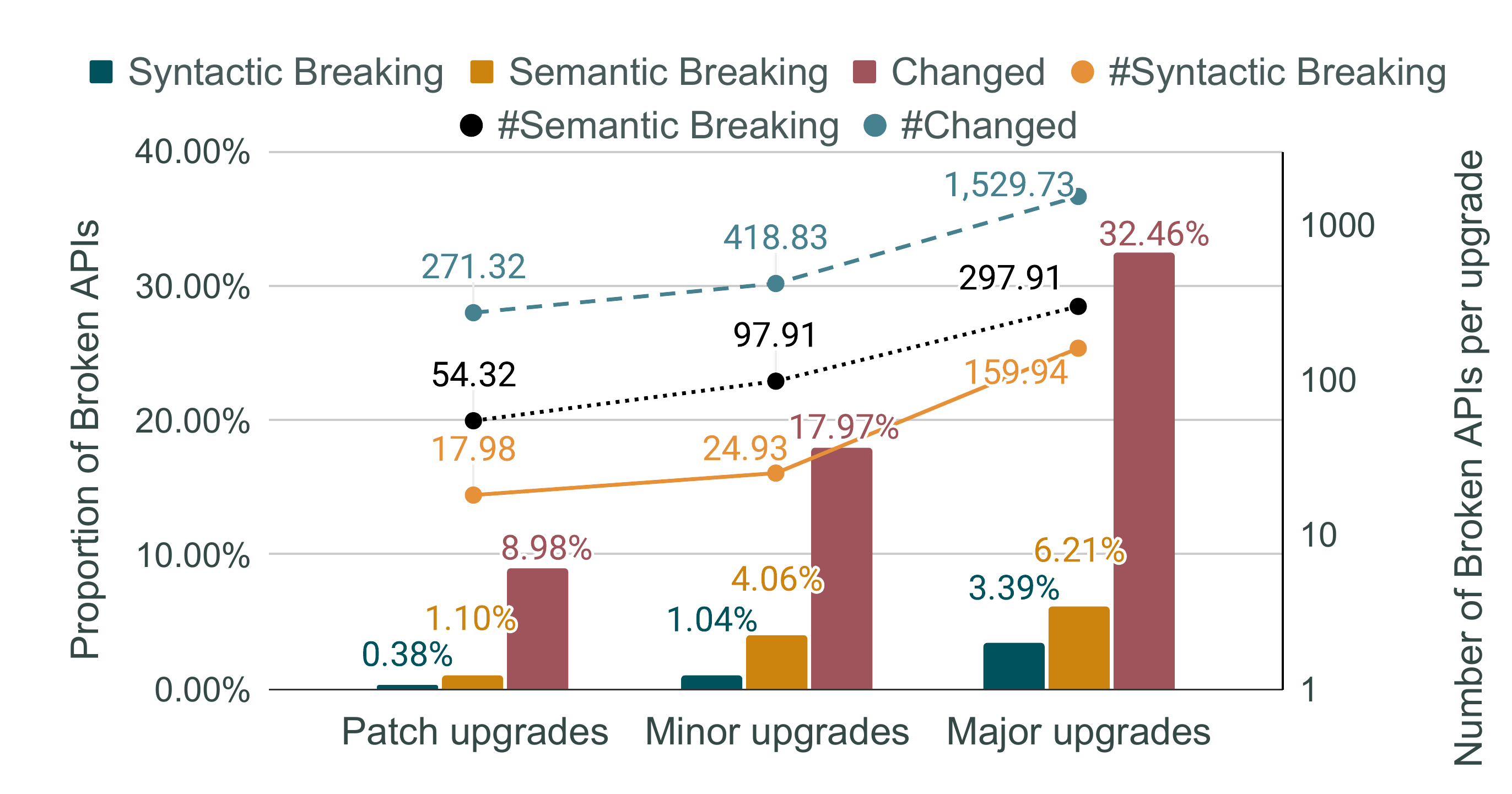}
  \caption{Average number and percentage of APIs over three upgrades}
  \label{fig:rq4_2}
\end{figure}

Considering version release strategies vary greatly among libraries, SemVer rules are not generally followed by popular libraries. Apart from \synb, \semb is also prevalent over \pnm upgrades. For developers, \pnm upgrades should check \semb too before publishing to avoid breaking downstream projects. For users, to avoid identifying version release strategies of dependencies case by case, \tool can be used on dependencies to pre-check the breaking APIs before upgrading along with \synb checking tools.

The \semb and \synb proportions at the API level are illustrated in Figure \ref{fig:rq4_2},  \textit{Changed} denotes the number of APIs with binary changes. It is observed that, for \semb, \tool significantly filtered out non-breaking APIs from \textit{Changed} APIs. The \textit{\synb}  APIs were way fewer than \textit{\semb}, which means \textit{\synb} checking is far from enough. Moreover, there were still $1.10\%$ APIs affected by the \semb in \textit{patch} upgrades and $4.06\%$ in \textit{minor} upgrades.

Since these libraries were dependencies of over 110k artifacts (libraries) in the Maven ecosystem, the unexposed \semb APIs could unexpectedly detriment the functionalities of those artifacts.

\begin{boxedtest}
\textbf{Conclusion of RQ3: }
From the experiment with $1,629,589$ APIs in $546$ version pairs, $1.10\%$ of APIs from \textit{Patch} and $4.06\%$ of APIs from \textit{Minor} upgrades were affected by \semb. They are 2-4 times more than \synb APIs ($0.38\%$ and $1.04\%$). In terms of the $497$ \pnm version pairs, \pa upgrades have $33.83\%$ breaking pairs, and \mi upgrades have $64.42\%$ breaking pairs because version release strategies adopted by libraries vary greatly.
\end{boxedtest}

\section{Threats to Validity}\label{sec:threats}
The primary threat is that benign behavior filtering cannot ideally reflect the real intention of the developers, because the rules to filter out harmless behaviors were made based on the empirical summary. The commit intention classification technology can be used to facilitate the accuracy of benign change identification, but they would introduce heavy procedures, such as Machine Learning models, which handicap the scalable and efficient deployment.

Another threat is the scope of \textit{API}. We took public methods of instantiable classes as APIs which are a superset of client-used APIs. However, since Java has no built-in indicator to mark exposed APIs, it is hard to accurately locate the actually used APIs. Taking advantage of usage data of TPL methods is plausible, but it is still not accurate.
\balance

\section{Related Work}\label{sec:related}
\noindent
\subsection{Study of Semantic Versioning Compliance}
Many research works \cite{lam2020putting,raemaekers2014semantic,raemaekers2017semantic,decan2019package,liu2022demystifying,ochoa2021breaking,abdalkareem2022machine} have studied the compliance of SemVer since its release. Raemaekers et al. \cite{raemaekers2014semantic, raemaekers2017semantic} found around $1/3$ of all releases introduce at least one breaking change in seven years release history of Maven Central Repository. Decan et al. \cite{decan2019package} studied 4 ecosystems (Cargo, NPM, Packagist, and Rubygems) to understand to what extent developers rely on SemVer to determine dependency constraints and found situations varying greatly among them. Ochoa et al. \cite{ochoa2021breaking} revealed that $83.4\%$ of upgrades of Maven comply with SemVer rules, and most breaking changes do not affect clients with only $7.9\%$ of clients affected. The works drew conclusions based on signature-based incompatibility instead of \semb, which is not complete. Thus, \tool is required to provide a more comprehensive analysis of the disobeying of SemVer by including \semb into the picture.

\subsection{API Compatibility Checking}
Only a limited number of research works \cite{mostafa2017experience,chen2020taming} regarding \semb of Java program were published in recent years. 
Mostafa et al. \cite{mostafa2017experience} conducted an empirical study on behavioral incompatibility phenomena in popular Java libraries and analyzed published issues from Jira. DeBBI \cite{chen2020taming} used cross-project testing to amplify the testing coverage to detect Behavioral Incompatibility. Their implementations relied on unit tests, thus subject to coverage. But \tool relies on static analysis so that \tool can conduct a more comprehensive analysis.
Towards analyzing or detecting the signature-based API compatibility issues of Maven or Android programs, massive empirical studies~\cite{xia2020android, huang2018understanding, jezek2015java, jezek2017api, xavier2017historical, wang2020empirical, dietrich2016java,contractcom}
have been conducted. RAPID \cite{xia2020android} detects the status of incompatible APIs in the Android ecosystem. Huang et al. \cite{huang2018understanding} studied callback compatibility issues of Android  and developed a tool based on CFG to detect such issues. Jezek et al. \cite{jezek2017api} evaluated 9 commonly used syntactical incompatible API detection tools. Apidiff \cite{brito2018apidiff} determined the incompatibility at the name level of methods used in the target library. CiD \cite{li2018cid} tried to alert the users by modeling the life cycle of APIs used in specific versions, while ACRYL \cite{scalabrino2019data} was a complementary method for CiD based on an alternative data-driven approach . 
The studies and detection tools based on the signature of APIs did not entail the semantics, thus, they cannot be used to detect \semb APIs like \tool.

\subsection{Software Evolution Studies}
Many works \cite{nakakoji2002evolution,bogart2016break,hora2015developers,hora2014apievolutionminer,dig2006apis,mcdonnell2013empirical,espinha2014web,bavota2015apache,bavota2013evolution,wu2016exploratory,wu2010aura,wu2015impact} were dedicated to studying the evolution of software across platforms. 
Dig et al. \cite{dig2006apis} discovered that $80\%$ of breaking changes belonged to refactoring over the evolution. Mcdonne et al. \cite{mcdonnell2013empirical} discovered that the adoption of API is much slower than the API evolution. Researchers of \cite{espinha2014web,bavota2015apache} summarized the best practice for developing web application APIs. Bavota et al. \cite{bavota2013evolution} revealed the impact of dependencies upgrade based on 14 years of published maven projects. Wu et al. \cite{wu2016exploratory,wu2010aura,wu2015impact} found that APIs in frameworks are more susceptible to the missing method or class. These studies established the foundations of API compatibility. With their contributions to the overall understanding of API compatibility, we can locate and resolve the pain points of API compatibility.

\section{Conclusion}\label{sec:conclusion}
We proposed \tool to {statically} detect \semb based on APIs during \pnm upgrades to enhance the compliance of SemVer rules. 
Experimental results demonstrated that \tool achieved $90.26\%$ recall and $81.29\%$ precision. Another experiment proves that \tool with larger coverage detected 4.5 times more APIs than the commonly used solution, unit tests. Furthermore, a study was conducted on the top 21 Java libraries for over 1.6 million APIs, and 546 version pairs to evaluate the compliance with SemVer rules at the library, version pair, and API levels, which revealed that $33.83\%$ \pa upgrades and $64.42\%$ \mi upgrades had at least one API affected by any breaking. And on average, there were 2-4 times more APIs affected by \semb issues than \synb issues.

\begin{acks}
This research is partially supported by the National Research Foundation, Singapore under its the AI Singapore Programme (AISG2-RP-2020-019), the National Research Foundation, Prime Ministers Office, Singapore under its National Cybersecurity R\&D Program (Award No. NRF2018NCR-NCR005-0001), NRF Investigatorship NRF-NRFI06-2020-0001, the National Research Foundation through its National Satellite of Excellence in Trustworthy Software Systems (NSOE-TSS) project under the National Cybersecurity R\&D (NCR) Grant award no. NRF2018NCR-NSOE003-0001, the Ministry of Education, Singapore under its Academic Research Fund Tier 2 (MOE-T2EP20120-0004) and Tier 3 (MOET32020-0004). Any opinions, findings and conclusions or recommendations expressed in this material are those of the author(s) and do not reflect the views of the Ministry of Education, Singapore.
\end{acks}

\clearpage

\bibliographystyle{ACM-Reference-Format}
\bibliography{acmart}

\end{document}